\documentclass[prl,english,aps,superscriptaddress,twocolumn,floatfix]{revtex4-1}

\usepackage{epsf}
\usepackage{graphicx}
\usepackage{sidecap}

\usepackage{amsmath}
\usepackage{amssymb}
\usepackage{bm}

\usepackage{color}
\usepackage{ulem}

\def \beq {\begin{equation}}
\def \eeq {\end{equation}}
\begin{document}

\title{{Microscopic Theory of Ultrafast Out-of-Equilibrium Dynamics in Magnetic Insulators. Unraveling the Magnon-Phonon Coupling}}

\author{Pablo Maldonado}
\affiliation{Department of Physics and Astronomy, Uppsala University, P.\,O.\ Box 516, S-75120 Uppsala, Sweden}
\author{Yaroslav O. Kvashnin}
\affiliation{Department of Physics and Astronomy, Uppsala University, P.\,O.\ Box 516, S-75120 Uppsala, Sweden}

\begin{abstract}
\noindent
The interaction between lattice and spins is at the heart of an extremely intriguing ultrafast dynamics in magnetic materials. In this work we formulate a general non-equilibrium theory that disentangles the complex interplay between them in a THz laser-excited antiferromagnetic insulator.
The theory provides a quantitative description of the transient energy flow between the spin and lattice sub-systems, subject to magnon-phonon and phonon-phonon scatterings, giving rise to finite life-times of the quasiparticles and to the equilibration time of the system.
We predict a novel kind of scattering process where two magnons of opposite polarizations decay into a phonon, previously omitted in the literature.
The theory is combined with first-principle calculations and then applied to simulate a realistic dynamics of NiO.
The main relaxation channels and hot spots in the reciprocal space, giving the strongest contribution to the energy transfer between phonons and magnons are identified. The diverse interaction strengths lead to distinct coupled dynamics of the lattice and spin systems and subsequently to different equilibration timescales.
\end{abstract}
\date{\today}
\maketitle
\textit{Introduction.}
The discovery of laser-induced subpicosecond demagnetization of metallic Ni by Beaurepaire \textit{et al.} was the first demonstration of the reduction of the magnetic order in a ferromagnetic system, happening several order of magnitude faster than the, at that time assumed, limiting timescale of a ferromagnetic demagnetization\cite{PhysRevLett.76.4250}. This finding triggered the dawn of ultrafast magnetism, that fueled by further experimental works \cite{RevModPhys.82.2731} has led to the discovery of new physical phenomena, such as ultrafast coherent control of spin waves \cite{NatPhys.5.31}, ultrafast generation of ferromagnetic order \cite{PhysRevLett.93.197403} and all-optical magnetic switching \cite{PhysRevLett.99.047601,Radu2011,Stupakiewicz2017}. Additionally, this research field is at the heart of novel technological applications, such as heat-assisted magnetic recording, and of the foundation of the fast development of ultrafast spintronic science \cite{0022-3727-50-36-363001}. Further, the use of ultrafast THz laser pulses has recently made it possible to gain access to fundamental low-energy excitations in a material, and thus experimentally monitor the energy and angular momentum transfer between lattice vibrations and spin waves \cite{Hashimoto2017,Maehrleineaar5164}. These experimental observations shine light on the coupling between phonon and magnons, which is a cornerstone in understanding not only ultrafast magnetization dynamics but also other effects such as magnetostriction, spin damping in magnetic insulators or magnon transport \cite{Chumak2015,Reid2018}. However, the microscopic description of this coupling and its relevance on those effects remain elusive and is one of the central challenges of contemporary condensed matter physics \cite{PhysRevB.96.100406,Holanda2018}.

Although several theoretical descriptions have been proposed to explain the ultrafast demagnetization of metals, they rely either on the direct interaction between the laser field and the magnetic system \cite{PhysRevLett.117.137203}, on the spin-orbit dependent electron-lattice coupling \cite{PhysRevLett.95.267207} or on superdiffusive spin transport \cite{PhysRevLett.105.027203}. Therefore, these models lack an explicit coupling between spin and lattice, which is only present in the phenomenological three-temperature model (3TM) \cite{PhysRevLett.76.4250}. Despite the 3TM is extensively used to explain the experimental findings, it lacks a microscopic derivation and should only be used under the assumptions of thermalized spin, lattice and electron subsystems, which are not fulfilled in the regime of ultrafast dynamics\cite{NatPhys.7}. On the other hand, various improved models have recently been developed aiming to disentangle the complex coupling between spin and lattice. While they incorporate a specific form of the spin-lattice coupling, the scope of those works lies outside of ultrafast out-of-equilibrium regime. Thus, these models either do not provide any solution to the out-of-equilibrium dynamics \cite{PhysRevB.89.184413,PhysRevMaterials.1.074404} or they address spin-lattice dynamics under classical considerations in equilibrium \cite{PhysRevMaterials.2.064401,2018arXiv180403119H}. Additionally, in the latter models the spin-lattice coupling is only derived for specific lattice vibrations -phonon eigenmodes- which, as we will see later, results in an incomplete description of the coupling mechanism.

Studying the ultrafast dynamics of magnetic insulators is of particular interest. 
The peculiarity of such systems lies in the fact that electron degrees of freedom are frozen and do not contribute to the system's dynamics.
Since electronic fluctuations are the fastest, one would expect a much slower spin dynamics in insulators as compared to metals.
However, recent results for YIG have shown that the magnetic order in an insulator can be altered on a picosecond time scale, which is much faster than previously expected \cite{Maehrleineaar5164}.
The microscopic mechanism responsible for this effect is currently not fully understood and it is something we address in this work.

\begin{figure}[th]
\vspace*{-0.2cm}
\includegraphics[width=0.99\linewidth]{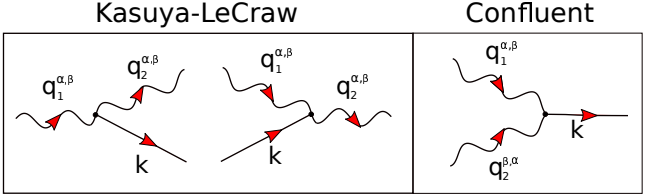}
\caption{Schematic diagrams showing the relevant magnon-phonon scattering processes.}
\label{Fig1}
\end{figure}

Here, we present a novel and general theory that dictates the laser-induced out-of-equilibrium system dynamics, and governs the energy transfer between lattice (phonons) and magnetic (magnons) subsystems in a magnetic insulator. 
The main features of the theory are an explicit inclusion of the anharmonic effects and a realistic branch and wavevector dependent description of the coupling between phonons and magnons. These couplings give rise to the quasiparticles lifetimes and to a balanced energy flow between them when excited. Strikingly, the theory shows that the system dynamics is not only driven by the known Kasuya-LeCraw (or Cherenkov) process in which one magnon decays into another magnon by emitting or absorbing a phonon, but also by a confluent process, in which two magnons of opposite polarization decay into a phonon (see Fig.~\ref{Fig1}).

Complemented by first-principles calculations, the theory acquires unprecedented predictive power and can be used to systematically address real material dynamics.
We use NiO as a benchmark material to test our out-of-equilibrium model obtaining not only the system relaxation dynamics and the phonon-magnon lifetimes, but also identifying the channels and the regions of the reciprocal space with the largest magnon-phonon (mg-ph) and phonon-phonon (ph-ph) interaction strengths.

\textit{Theory and methods.}
To describe the non-equilibrium time evolution of the magnonic and phononic degrees of freedom of the laser-excited material we follow an analogous description to the model proposed by Maldonado \textit{et al.} \cite{PhysRevB.96.174439}, applied in our case to a magnetic insulator. Specifically, we describe the lattice and magnetic systems by dividing them into $N$ independent phonon and magnon subsystems, respectively, each of them characterized by a specific branch $\nu$, and a wavevector $\bm{q}$ ($Q\equiv (\bm{q},\nu)$). The phonons interact with one another through ph-ph scattering and with the magnons via mg-ph scattering. These interactions are wavevector and branch dependent. Therefore, the out-of-equilibrium phonon and magnon subsystem populations, $N_Q$ and $n_{Q}$, evolve separately but not independently in time. 
The model further assumes that an ultrashort THz laser pulse excites the infrared phonon modes in a small region around the $\Gamma$ point of the Brillouin Zone (BZ). Subsequently, the excited phonon modes relax transferring energy into other phonon and magnons modes via ph-ph and mg-ph couplings.

A theoretical formulation can be achieved by making use of the classical kinetic theory and total energy conservation. The total spin or lattice energy is given by  $E =\sum_Q\hbar\omega_Q n_Q$, where  $\hbar \omega_Q$ is the quasiparticle energy and $n_Q$ the out-of-equilibrium populations. For the latter we use the following ansatz for the out-of-equilibrium lattice and spin populations:

\begin{equation}
n_Q (\Psi_{Q}) \!=\!{\Big[{e^{\frac{\hbar\omega_{Q}}{k_{B} T}+\Psi_{Q}(t)}-1} \Big]^{-1}},
\end{equation}

where $\Psi_{Q}(t)$ is a measure of the deviation from the equilibrium Bose-Einstein distribution for phonons or magnons in the mode $Q$ \cite{srivastava1990physics}. 
The short time scale of the out-of-equilibrium microscopic processes involved leads to a simplification of the kinetic equations where the diffusion term can be neglected. Thus, we can define the energy exchange rates as

\begin{align}
\frac{\partial E_{\ell}}{\partial t}=&\sum_Q \hbar\omega_Q \frac{\partial N_Q}{\partial t} \big|_{\text{mg-ph}}^{scatt.}+\sum_Q \hbar\omega_Q \frac{\partial N_Q}{\partial t} \big|_{\text{ph-ph}}^{scatt.}+u(t),\label{rate-E1}\\
\frac{\partial E_{s}}{\partial t}=&\sum_{\sigma}\sum_Q \hbar\omega_Q \frac{\partial n^{\sigma}_Q}{\partial t} \big|_{\text{mg-ph}}^{scatt.},\label{rate-E2}
\end{align}

where the sums over $\sigma$ on the magnon population stands for the different magnetic polarizations and the subscripts denote the different scattering processes that change the distributions. The $u(t)$ is the source term, that represents the laser driving field. 
To determine the energy flow between the magnon and phonon subsystems we have to solve the Eqs.~(\ref{rate-E1}) and (\ref{rate-E2}), which govern the time evolution of the boson populations. These populations experience changes due to various scattering events, which can be derived from the Fermi's Golden rule of the scattering theory (see Supplemental Material (SM) \cite{SM} for explicit forms of these terms). 
Finally, we obtain a set of coupled differential equations which connect the energy rate of different phonon and magnon modes:

\begin{widetext}
\begin{align}
\displaystyle{\frac{d\epsilon^{\alpha}_{q}}{dt}=\hbar\omega^{\alpha}_q \left(\frac{dn^{\alpha}_{q}}{dt}\right)}=& \displaystyle{\hbar\omega^{\alpha}_q\sum_{\nu}\sum_{k} (\chi^{k,q}_1+\chi^{k,q}_2)(N_{k}(\Psi_{k})-N_{k}(\Psi_{q}))+u(t) ~~~~q=q_1, \ldots, q_N }\label{rate-T1} \\
\displaystyle{\frac{d\epsilon_{k}}{dt}=\hbar\omega_{k} \left(\frac{dN_{k}}{dt}\right)}=&\displaystyle{\hbar\omega_{k}\sum_{q} (\chi^{k,q}_1+2\chi^{k,q}_2)(N_{k}(\Psi_{k^{\prime}})-N_{k})+\hbar\omega_{k} \sum_{\nu}\sum\limits_{k^{\prime}} \Gamma_{kk^{\prime}} \left( n_{k^{\prime}}(\Psi_{k^{\prime}}) -n_{k}(\Psi_{k}) \right) ~~~~k=k_1, \ldots, k_N.}\label{rate-T2}
\end{align}
\end{widetext}

where $\epsilon^{\alpha}_{q}$ and $\omega^{\alpha}_q$ ($\epsilon_{q}$ and $\omega_q$) are the total and individual magnon (phonon) mode-dependent energies for each polarization $\alpha$, respectively. $\chi^{\bf{k,q}}_1$ and $\chi^{\bf{k,q}}_2$ are the magnon/phonon linewidths, originating from the mg-ph confluent and Kasuya-LeCraw processes. $\Gamma_{kk^{\prime}}$ is the phonon linewidth, caused by the ph-ph interaction. 
The $\chi^{\bf{k,q}}_1$ term naturally appears in the derivation presented the SM\cite{SM} and to our knowledge has not been discussed previously in the literature.

If the laser driving field, $u(t)$, resonantly excites the THz phonon modes, then this energy initiates an out-of-equilibrium dynamics of the system, which proceeds as follows.
The ph-ph coupling shares the injected excess of energy among other phonon modes, thus guiding the lattice towards equilibration. 
Contrarily, the mg-ph coupling conducts the energy from the lattice to the spins, bringing it into a non-equilibrium state with highly non-homogeneous distribution of energy. But at the later stage, it is this coupling that will initiate the spin equilibration. 
Hence, the characteristic feature of the model is this intrinsic competition between ph-ph and mg-ph interactions, which determines not only the different subsystems equilibration times, but also the out-of-equilibrium system dynamics and timescale of the process. 

Note that the here-proposed theory is general and can be applied to any kind of laser excitation and any specific form of the mg-ph interactions, such as exchange interaction or dipole and spin-orbit interactions.

\textit{Application to a real material.}
The model, defined by Eqs.~\eqref{rate-T1} and \eqref{rate-T2}, can be applied to study the ultrafast dynamics in real materials.
For this purpose, one has to compute the material-specific properties, such as the phonon and magnon energies and the mode-dependent magnon and phonon linewidths. 
We have performed \textit{ab initio} calculations of these quantities for NiO, which is an excellent material for antiferromagnetic spintronics applications\cite{RevModPhys.90.015005,Nemec2018,Satoh2014}.
We used density functional theory combined with Hubbard $U$ correction (DFT+$U$) and the details of the calculations can be found in SM \cite{SM} along with the relevant references \cite{phonopy,lsdau,ldau1,NiO-exp-mg,Jacobsson-TMOs,Savrasov-jijDMFT,keffer,kotani-magnons,jij-orig-1,PhysRevB.97.184404,rspt-book,jij-ldapp-2000,rspt-jij,j-tmo-2}. It is important to mention that the ph-ph interaction was treated using the many-body perturbation theory in a third-order anharmonic approximation\cite{PhysRevB.91.094306}.
The magnetic system of NiO was described by the Heisenberg model. 
We have extracted the exchange parameters and their gradients along the phonon eigenvectors for each phonon Q (for details, see SM \cite{SM}).
The latter are fundamental quantities for the determination of the $\chi_1^{\bf{k},\bf{q}}$ and $\chi_2^{\bf{k},\bf{q}}$ linewidths.
The laser driving field, $u(r)$, was considered to be a 100-fs (FWHM) gaussian pulse of the energy $\omega=11.5$ THz that excites resonantly the corresponding infrared phonon modes.

\begin{figure}[th]
\vspace*{-0.2cm}
\includegraphics[width=0.99\linewidth]{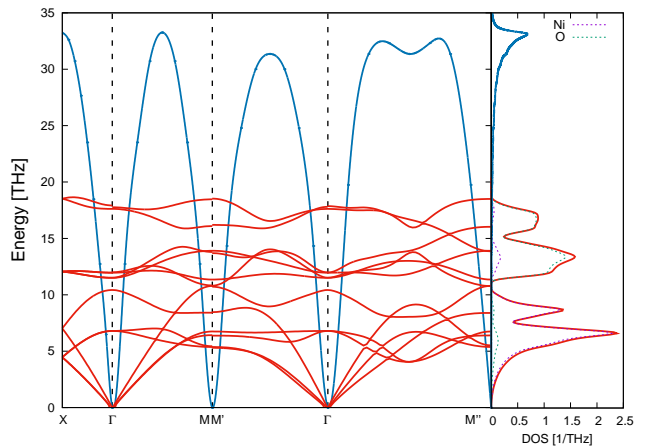}
\caption{(a) Calculated magnon (blue) and phonon (red) dispersions in antiferromagnetic NiO. (b) The corresponding magnon and phonon density of states (DOS). The phonon DOS is projected onto Ni and O atomic contributions. Note that there are two formula units in the magnetic unit cell and thus there are 12 phonon branches, which can be unfolded into the 6 branches of the primitive unit cell. The coordinates and the labeling of the high symmetry points follows the usual notation and are taken from Ref.~\cite{PhysRevB.88.134427}. 
}
\label{Fig2}
\end{figure}

\textit{Results.}
As seen in Eqs.~\eqref{rate-T1} and \eqref{rate-T2} the mode dependent phonon and magnon energies are have a direct relevance for the system's dynamics. 
In Fig.~\ref{Fig2} (a) we show our calculated magnon (in blue) and phonon (in red) dispersions along high-symmetry lines in the magnetic BZ. 
Fig.~\ref{Fig2} (b) shows the \textit{ab initio} computed magnon and projected and total phonon density of states (DOS). 
As expected, we find that the O and Ni atoms contribute mainly to high and low energy states respectively, with very small overlap between them. 
It is interesting to note that although phonon and magnon lines show multiple crossings at the energies between 5 and 20 THz, the overlap between their DOS is minimal, since the magnons are concentrated at higher energies (between 25 and 30 THz).
Based on these results, one can already anticipate the mg-ph scattering to be relatively suppressed in this system.

\begin{figure}[th]
\vspace*{-0.2cm}
\includegraphics[width=0.99\linewidth]{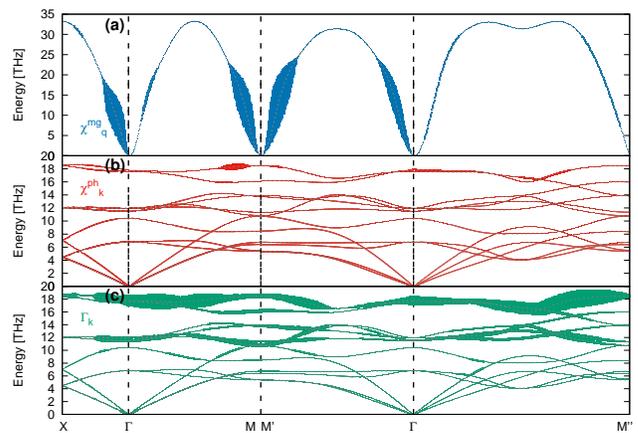}
\caption{Calculated magnon (a) and phonon linewidths due to mg-ph and ph-ph interactions, (b) and (c) respectively, for the different magnon and phonon branches along the high symmetry lines. The symbol size is proportional to the magnitude of the k-dependent mg-ph and ph-ph coupling functions at 300 K. Magnon and phonon dispersions are represented by the dashed lines.}
\label{Fig3}
\end{figure}

The total magnon ($\chi_q^{mg}=\frac{1}{N_k}\sum_k (\chi_1^{k,q}+\chi_2^{k,q})$) and phonon ($\chi_k^{ph}=\frac{1}{N_q}\sum_q (\chi_1^{k,q}+2\chi_2^{k,q})$) linewidths due to mg-ph interaction, and the total phonon linewidth ($\Gamma_k=\frac{1}{N_{k^{\prime}}}\sum_k\Gamma_{kk^{\prime}}$) due to ph-ph interaction are shown in Fig.~\ref{Fig3}, together with the magnon and phonon dispersions (dashed lines) along the high symmetry lines. 
The magnitude of the wavector dependent couplings are proportional to the symbol size. 
Remarkably, that although the matrix elements due to the ph-ph interaction are much smaller in magnitude than those due to mg-ph interaction, the linewidths provided by the two types of couplings are of the same magnitude, thus evidencing their similar coupling strengths.
To understand this striking result, it is necessary to analyze the phonon and magnon DOS and the mechanism of the mg-ph coupling. 
As illustrated in Fig.~\ref{Fig1} two different microscopic processes can couple phonons and magnons, i.e. confluent and Kasuya-LeCraw processes. 
In the confluent process two magnons decay into a phonon (or the opposite process) of the same energy. Since there are very few magnon states having lower energies than the phonon states, the probability of this process becomes highly suppressed, leading to a negligible contribution. 
On the other hand, the Kasuya-LeCraw process involves a phonon and a magnon to produce (or annihilate) another magnon. 
Although this process is also limited by the energy and momentum conservation, it is more effective due to a larger number of possible combinations. 
As a consequence, the magnon linewidths are highly anisotropic with a strong dependence on the the magnon energies and momenta, as seen in Fig.~\ref{Fig3} (a). It can also be seen that the magnon states with low energies contribute significantly more to the coupling. 
Similarly, phonon linewidths due to mg-ph coupling, illustrated in Fig.~\ref{Fig3} (b), also exhibit a strong branch and wavector dependence. 
Thus, the mg-ph interaction mainly couple magnons with high energy phonon states localized at very specific regions of the reciprocal space. 
The main contribution to the energy flow between phonons and magnons will come from the coupling of those phonon states with low-energy magnons with large linewidths (see Fig.~\ref{Fig3} (a)), while the contribution stemming from other phonon and magnon states will be negligible. In contrast, ph-ph interaction involves three phonon modes with energies distributed in the same energy window, leading to an increased number of coupling processes and therefore to a large phonon linewidths. This is illustrated in Fig.~\ref{Fig3} (c), where the strong dependence on the phonon linewidths on wavevectors and branches can also be seen. 

Before we proceed to study the ultrafast dynamics, it is important to mention two crucial points. 
First, the observed anisotropy of the phonon and magnon linewidths due to  mg-ph coupling originates partially from the high Q-dependence of the gradients of the $J_{ij}$'s.
For NiO, the optical phonon modes, mainly involving oxygen motion, were found to completely define the linewidths and the low-energy branches gave negligible contribution to it.
Thus, we would like to stress that assessing the mg-ph coupling assuming the displacement of individual magnetic ions only, as was done e.g. in Refs. \cite{PhysRevMaterials.2.064401,2018arXiv180403119H}, might be a too crude approximation for magnetic oxides.
Second, even if a certain collective displacement substantially modifies some exchange coupling, it does not necessary contribute to the mg-ph \textit{coupling}, due to the restrictions, imposed by the energy and momenta conservation laws.

\begin{figure}[th]
\vspace*{-0.2cm}
\includegraphics[width=0.95\linewidth]{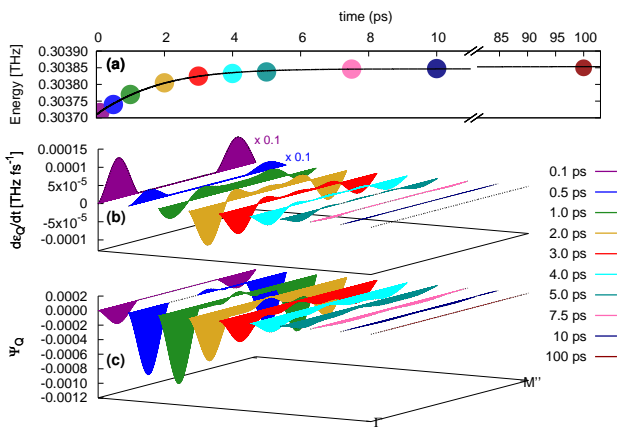}
\caption{(a) Energy of the spin system as a function of time. (b) Wavevector-dependent magnon energy flow (b) and deviation from equilibrium function (c) along the high symmetry line $\Gamma$-M$^{\prime\prime}$ at different time intervals. Energy rates at 0.1 and 0.5 ps have been multiplied by a factor 0.1 for better visualization.}
\label{Fig4}
\end{figure}

Using the \textit{ab initio} derived phonon and magnon dispersions and linewidths as input parameters, the set of coupled differential equations (Eq.~\eqref{rate-T1} and \eqref{rate-T2}) are solved numerically, and the out-of-equilibrium dynamics of the system is revealed. 
To illustrate this dynamics, in Fig.~\ref{Fig4}, we plot the energy of the spin system as a function of time (panel (a)) together with the Q-dependent energy flow ($\frac{d\epsilon_Q}{dt}$) and the $\Psi_Q$ function (panels (b) and (c), respectively). 
Fig.~\ref{Fig4} (a) shows that the increase of the spin energy starts immediately following the laser excitation, and it is only within the first 5 ps when most of the energy is delivered into the spin system. 
This is also evidences by Fig.~\ref{Fig4} (b), where $\frac{d\epsilon_Q}{dt}$ reaches its highest positive values only at very early times (0.2 ps). Afterwards, the flow of energy not only decreases, but even becomes negative at the parts of the BZ, where the mg-ph coupling is the largest (hot spots), which corresponds to the regions of the highest energy flow from the lattice. This negative rate reflects that those magnon states start redistributing their excess of energy among other magnon and phonon states in larger amount that the incoming energy from the lattice. We can therefore identify this time as the beginning of the spin equilibration. 

To illustrate the timescale of the equilibration, the measure of the deviation from equilibrium of the spin dynamics, $\Psi_Q$, is shown in Fig.~\ref{Fig4} (c).
Initially, and shortly after laser excitation the spin system is driven out of equilibrium by the mg-ph coupling with $\Psi_Q$ reaching its maximal values at around  1 ps, corresponding with the time at which the energy rate becomes negative at the hot spots of the BZ. The deviation is the largest at those regions due to the large amount of energy they have gained. However, when the spin equilibration is initiated, $\Psi_Q$ starts decreasing slowly converging towards zero (i.e. equilibrium) at long timescales (about 10 ps). Additionally, our results show that the lattice starts to equilibrate immediately after being irradiated and reaches quasi-equilibrium (i.e. 95$\%$ of the BZ deviate within 0.01 K -see SM \cite{SM}-) after 25 ps, while the spin system is driven out-of-equilibrium during the first picosecond, then starts to equilibrate and reaches the same degree of order at earlier times (9 ps).  

Hence, we have shown that THz excitation of infrared phonon states in NiO leads to two strongly coupled, but still distinct coupled out-of-equilibrium dynamics of the lattice and spin systems, due to the similar strengths of the interaction mechanisms. 
On one hand, the lattice is rapidly brought out of equilibrium excited by the radiation and starts to equilibrate immediately after. On the other hand, the spin system is driven by phonon excitation and only starts to equilibrate after 1 ps. The difference is that the laser radiation excites a very specific phonon mode, resulting in a highly inhomogeneous state of the lattice, whereas the spin system, subjected to the mg-ph coupling, gets excited more uniformly. As a result, the equilibration time for the latter is shorter.

\textit{Conclusions.}
We have formulated a general non-equilibrium theory that disentangles the complex microscopic interplay between lattice and spins. This novel theory not only produces a comprehensive solution of the laser-induced out-of-equilibrium dynamics, but also determines the equilibration times of the lattice and spin systems. Additionally, the theory provides the specific channels of the transient energy flow that drives the system dynamics, subject to mg-ph and ph-ph interactions. Such interaction defines another relevant feature of the model, which is the microscopic determination of the phonon and magnon linewidths. They identify the main relaxation channels and hot spots in the reciprocal space, giving the strongest contribution to the energy transfer between phonons and magnons. The theory also predicts the presence of a novel confluent scattering process between phonon and magnons due to the exchange interaction that was assumed hitherto forbidden. 

Furthermore, the combination of the theory with first principle calculations allows the model to quantitatively solve the realistic dynamics of a laser-excited system. The reliability of the model has been tested by studying the dynamics triggered by resonant THz excitation of phonon modes in NiO, where notably we discover that the mg-ph coupling has a strong Q-dependence. The interaction strength is of the same magnitude as the one produced by the ph-ph interaction, leading to a coupled, but still diverse dynamics of the lattice and spin systems. Thus, while the latter is driven out-of-equilibrium subsequent to the phonon excitation and has shorter equilibration times, the former starts equilibrating following the laser excitation and  reaches a steady state at larger times.  

The present theory, being general and parameter-free, is expected to become an important tool for describing not only laser-induced spin dynamics but also other ultrafast phenomena related with spintronics and magnetic magnetic storage.

\textit{Acknowledgments.} This work has been funded through the Swedish
Research Council (VR, Grant No. 2016-03875). We also acknowledge support from the Swedish National
Infrastructure for Computing (SNIC).

\bibliography{references} 

\end{document}